\documentclass[sigconf]{acmart}

\usepackage{booktabs} 
\usepackage{balance}
\setcopyright{none}
\setcopyright{acmcopyright}
\begin{document}

\copyrightyear{2018}
\acmYear{2018}
\setcopyright{acmcopyright}
\acmConference[SEEM'18]{SEEM'18:IEEE/ACM International Workshop on
Software Engineering Education for Millennials }{May 27-June 3,
2018}{Gothenburg, Sweden}
\acmBooktitle{SEEM'18: SEEM'18:IEEE/ACM International Workshop on
Software Engineering Education for Millennials , May 27-June 3, 2018,
Gothenburg, Sweden}
\acmPrice{15.00}
\acmDOI{10.1145/3194779.3194791}
\acmISBN{978-1-4503-5750-0/18/05}

\title{Teaching Requirements Engineering Concepts using Case-Based Learning}

\author{Saurabh Tiwari}\additionalaffiliation{M\"{a}lardalen University, Sweden}
\affiliation{DA-IICT, Gandhinagar, India}
\email{saurabh\_t@daiict.ac.in}

\author{Deepti Ameta}
\affiliation{DA-IICT, Gandhinagar, India}
\email{deepti\_ameta@daiict.ac.in}

\author{Paramvir Singh}
\affiliation{NIT Jalandhar, India}
\email{singhpv@nitj.ac.in}

\author{Ashish Sureka}
\affiliation{Ashoka University, India}
\email{ashish.sureka@ashoka.edu.in}

\begin{abstract}

Requirements Engineering (RE) is known to be critical for the success of software projects, and hence forms an important part of any Software Engineering (SE) education curriculum offered at tertiary level. 
In this paper, we report the results of an exploratory pilot study conducted to assess the effectiveness of Case-Based Learning (CBL) methodology in facilitating the learning of several RE concepts. The evaluation was made on the basis of graduate students' responses to a set of questions representing various key learning principles, collected after the execution of two CBL sessions at DA-IICT, Gandhinagar (India). We investigate the perceived effectiveness of CBL in students' learning of various RE concepts, based on factors like case difference, gender diversity, and team size. Additionally, we collect and analyze the Teaching Assistants' (TAs) opinions about the conducted CBL sessions. The outcome of this CBL exercise was positive as maximum students were able to achieve all the five stated learning objectives. The authors also report various challenges, recommendations, and lessons learned while experiencing CBL sessions. 

\end{abstract}


\maketitle

\section{Introduction}
Software requirements engineering, as a part of any conventional (e.g. Waterfall) or modern (e.g. Agile) software development models, encapsulates an important set of activities that are required to express the purpose and functionality of a software system. Gathering, understanding and analyzing a set of software requirements require systematic, quantifiable, and repeatable techniques that ensure completeness, consistency, and relevance of requirements~\cite{kotonya1998requirements}. RE is not an exact science and multiple alternative approaches and solutions are possible~\cite{hofmann2001requirements}. Hence, RE analysis can be best performed through discussion, brainstorming, critical thinking and analyzing problem domain from multiple perspectives~\cite{4685657}\cite{wieringa2004requirements}. 

RE teaching relies on an approach to convey RE related concepts to the students in a way that drives them towards analyzing the problem statement with multiple point-of-views and search for the best suitable solution~\cite{4685657}. Further, it is well known that educators face obstacles in teaching RE concepts due to its multidisciplinary nature which deals with both computer science and social sciences concepts~\cite{Mead:2006}. In recent years, the use of active learning (different than the traditional lecture-based approaches) approaches for teaching RE concepts has gained a lot of interest among software engineering educators, as these approaches involve discussions, teamwork, decision-making tasks, brainstorming, engagements, and critical thinking~\cite{Peng:2010}\cite{Zowghi:2003}. 

The CBL methodology~\cite{kundra2016}\cite{Papadopoulos:2007}\cite{veena2017} is also committed to achieving similar objectives, and can be used to teach some of the selected RE concepts. A case in CBL is a unique, complex, and uncertain narrative structure of some contemporary interest arousing event or problem~\cite{Papadopoulos:2007}. The existing literature on various teaching methodologies reflects that CBL has been used since long in the fields of Health Science education~\cite{williams2005case}, Law education, and Business education~\cite{Mahgoub:2012}. However, the application of CBL in teaching concepts of RE is unexplored. Looking at the importance of CBL~\cite{Papadopoulos:2007} and the issues of RE education~\cite{Ouhbi2015}\cite{portugal16}, we intend to introduce CBL for teaching various RE concepts in a graduate SE course. 

The main objective of this study is to examine the perceived effectiveness of CBL in teaching RE discipline. We develop two RE cases, conducted classroom CBL sessions, write experience reports, and share them publicly through Software Engineering Case-Based Learning Database (SEABED)\footnote{http://seabed.in/}. This paper makes four novel research contributions:
\begin{enumerate}
  \item First implementation of CBL for teaching and practicing the concepts of RE discipline for a large class of 112 students at DA-IICT, Gandhinagar (India)\footnote{http://www.daiict.ac.in/}.
  \item Empirical analysis based on case difference, gender diversity, team size, and engagement between the organizing team and the students during CBL sessions. 
  \item A set of challenges and recommendations derived from the experiences gained while implementing two CBL sessions, and strengthened by the experiences shared by the authors from three different universities collaborating and teaching SE for several years.
  \item Two original RE cases, \emph{Metro Ticket Distributor System}\footnote{http://seabed.in/case-study/Metro\_Case.pdf} and \emph{LIC Market-Driven System}\footnote{http://seabed.in/case-study/LIC\_Case.pdf}, available on SEABED.
\end{enumerate}


\section{Related Work}
Teaching RE concepts using an active learning approach has gained a lot of interest in RE domain, and hence several papers (e.g.,~\cite{berenbach2005hole}\cite{Peng:2010} \cite{Ouhbi2015}\cite{Zowghi:2003}) about RE education have been published in literature. The main challenge is providing students with a logical understanding of the RE phase, and how the concepts they learned have been applied in real projects. Ouhbi et al.~\cite{Ouhbi2015} performed a systematic mapping study on RE education to classify the existing studies based on the research type, empirical type, contribution type, RE activity, and course curricula. The mapping study resulted in a list of advice obtained from the RE education literature for instructors. This list of advice covers: (1) teaching how to define the problem scope, and avoid general/vague specifications, (2) guiding how to select and use RE tools, (3) describing activities in requirements analysis and modeling, (4) involving students in industrial projects, and (5) familiarizing students with approaches to problem-solving and development methodologies.

Portugal et al.~\cite{portugal16} presented an experience report regarding challenges faced while teaching RE to undergraduate students. The study was conducted in three consecutive semesters with a total of 57 students by taking traditional lectures, applying project-based learning methodology, and collecting feedbacks. They suggested that various RE concepts such as project planning, quality control, client involvement, and budgeting can be taught using Project-Based Learning (PBL). PBL is one of the most commonly used teaching methodologies in SE education, which has its own set of limitations ~\cite{kundra2016}\cite{portugal16}. 

Another technique that gains a lot of interest of instructors is role playing in teaching RE concepts. Zowghi et al.~\cite{Zowghi:2003} conducted a study to teach RE through role-playing. The authors focused on various RE tasks like elicitation, analysis, modeling, validation, specification, and management. They concluded that the role-playing tool developed for teaching RE in the problem-solving mode, gives a better understanding of multiple perspectives on RE and the techniques employed to execute each underlying RE task. Peng et al.~\cite{Peng:2010} also applied a role-playing tool for teaching RE concepts. The results suggested that role playing approach is useful to incorporate the bidding of projects such that the interest or commitment of the developers for the project is increased. Svensson et al.~\cite{Svensson2017} investigated whether a role-playing project impacts students' scores against a written RE course examination. The results show that the students who received higher grades in the role playing project scored significantly higher as compared to the students with a lower project grades.

CBL is a teaching methodology that motivates students to read, understand, and discuss complex real-life scenarios, testing their analytical thinking and decision-making skills~\cite{kantar2015case}\cite{Peplow1997}. Garg et al.~\cite{garg2015} developed a case related to software architecture and introduced a Case-Oriented Learning Environment (COSEEd) for teaching Software Engineering concepts to undergraduate and graduate students. They found that COSEEd helps students to learn software engineering principles more efficiently than lecture-based learning. Saini et al.~\cite{veena2017} proposed an open source web-based Software Engineering Case-Based Learning Platform called SEABED. They conducted an experimental study to show the effectiveness of CBL methodology on students' learning, and also provided guidelines to write cases for CBL. 

Kundra et al.~\cite{kundra2016} used CBL for teaching the concepts of Compiler Design course where the authors reported their experiences in implementing case-based and project-based learning for teaching various Compiler Design concepts. Their results suggested that the case-based teaching enhances students skills of learning, critical thinking, engagement, communication skills and teamwork. In this paper, we focus on teaching and practicing various concepts of RE discipline using a CBL exercise, and investigate whether the use of CBL enhanced students' learning.

\section{Exploratory Pilot Study}
In this section, we present various elements of CBL implementation for teaching RE concepts as a part of a graduate-level SE course.

\subsection{Study Objectives and Aims (AIMs)}
In this work, we aim to investigate the effectiveness of CBL methodology in teaching/learning various RE concepts as a part of a SE course, by examining the achievement of a set of five students' learning objectives through empirically analyzing students' responses to CBL execution. These five learning objectives include students' learning, critical thinking, engagement, communication skills and teamwork. Accordingly, we frame a set of five objectives or aims (AIMs) which are stated as questions to address our proposed goal: \\

\noindent \textbf{AIM1:} \textit{Is CBL method effective in achieving various learning objectives?}

\noindent\textbf{AIM2:} \textit{Do students who worked in smaller groups show different responses to those who worked in larger groups?}

\noindent \textbf{AIM3:} \textit{Does CBL effectiveness in teaching RE concepts differ across two RE cases?}

\noindent \textbf{AIM4:} \textit{Is CBL effectiveness in teaching RE concepts influenced by gender diversity?}

\noindent \textbf{AIM5:} \textit{Does CBL result in a better engagement between TAs and students?}

\subsection{Subjects}
The subjects were 112 second year postgraduate students of MSc (IT) studying a compulsory course on \emph{IT 632 Software Engineering} in Autumn 2017 at DA-IICT, Gandhinagar (India). The CBL exercise was conducted as a graded exercise for the course. The students were of the same age group between 20 to 23, and can represent Millennials or Generation X.


\subsection{Cases}
We created two RE cases each accompanied with a set of questions to be solved during the CBL exercises. The cases used are available online$^{3,4}$. The idea and elaboration of the two cases was proposed by the two primary authors, and then reviewed by the other two authors with a point-of-view of filling the practical learning gaps between academia and industry. The other two authors of the paper have a significant amount of experience in the academia/industry. Both the cases intend to facilitate the concepts of understanding the problem domain, requirement elicitation and prioritization through the real-world scenarios. Specifically, the RE concepts covered in the cases are requirement elicitation techniques, requirement analysis, requirement prioritization techniques, concept mapping, use cases and user stories.

\subsection{Exploratory Pilot Study Design}
We used the single-factor incomplete block design for our study~\cite{Wohlin2012}, where every student group did not work on both the cases, and the students' experiences is considered as a blocking factor. The effectiveness of CBL mainly depends on the sizes of the teams formed to discuss and identify appropriate case solutions, hence we evaluated the impact of group size on students' learning (AIM2). Also, maintaining homogeneity among the groups is important, and hence we created the groups based on the Cumulative Percentage Index (CPI) (or grade points) attained by students in their previous semesters such that the average CPI across the groups was comparable.

Since the female presence in computing domain is sparse, their participation and contribution are important for the society to have a more balanced and equal representation~\cite{7914585}\cite{goode2006lost}, we did this gender specific analysis to provide some useful insights in this regard (AIM4). Also, while dividing the students in different group sizes, we tried to maintain male and female ratios among the groups. The class size of 112 students was divided into 14 groups -- 8 groups of 6 students, 4 groups of 11 students and 2 groups of 10 students. The students were divided into a total of 14 different groups based on their CPI and gender. The details of the students groups created for CBL sessions are shown in Table~\ref{tab:table7}.

\begin{table}[t!]
\centering
\small
\caption{\textsc{Percentage of male and female students in each group}}
\vspace{-2.5ex}
\begin{tabular}{|c|c|c|c|c|} \hline
\textbf{Group} & \textbf{\# Members} & \textbf{\% Male} & \textbf{\% Female} & \textbf{Average CPI} \\ \hline
\textbf{G1} & 6     & 66 (\#4) & 34 (\#2) & 7.436 \\ \hline
\textbf{G2} & 6     & 50 (\#3) & 50 (\#3) & 7.478 \\ \hline
\textbf{G3} & 11    & 72 (\#8) & 28 (\#3) & 7.061 \\ \hline
\textbf{G4} & 10    & 70 (\#7) & 30 (\#3) & 7.257 \\ \hline
\textbf{G5} & 11    & 72 (\#8) & 28 (\#3) & 7.058 \\ \hline
\textbf{G6} & 6     & 66 (\#4) & 34 (\#2) & 7.060 \\  \hline
\textbf{G7} & 6     & 66 (\#4) & 34 (\#2) & 7.166 \\ \hline
\textbf{G8} & 6     & 66 (\#4) & 34 (\#2) & 7.116 \\ \hline
\textbf{G9} & 11    & 72 (\#8) & 28 (\#3) & 7.068 \\ \hline
\textbf{G10} & 11   & 72 (\#8) & 28 (\#3) & 7.064 \\ \hline
\textbf{G11} & 6    & 50 (\#3) & 50 (\#3) & 7.366 \\ \hline
\textbf{G12} & 10   & 70 (\#7) & 30 (\#3) & 7.062 \\ \hline
\textbf{G13} & 6    & 66 (\#4) & 34 (\#2) & 7.064 \\ \hline
\textbf{G14} & 6    & 66 (\#4) & 34 (\#2) & 7.257 \\ \hline
\end{tabular}
\label{tab:table7}
\vspace{-3.4ex}
\end{table}

Various characteristics of our study design with respect to each CBL session can be inferred from Table~\ref{design:table}. In our study design, we chose different sizes of students groups in order to analyze the impact of group size on student learning, critical thinking, and engagement of the concepts. Out of 112 students, 76 (68\%) were boys and 36 (32\%) were girls. While assigning students to each team, we ensure that the group must be balanced with respect to gender, and their CPI attained in their previous academic sessions. We evaluated students solutions for each of CBL exercises from their case presentations and case reports. We identified the best group for each case, and rewarded it with 10 marks in the final grading. A total time period of seven days was given to each group for solving the case. The various elements of our study are as follows:
\begin{itemize}
    \item \emph{Factor (Independent Variable)}: RE case
    \item \emph{Alternatives}: Case A (Metro Ticket Distributor System) and Case B (LIC Market-Driven System) 
    \item \emph{Response (Dependent) Variables}: Students responses with reference to case difference, gender diversity, team size
    \item \emph{Study Design Method}: Single-factor incomplete block design~\cite{Wohlin2012}.
\end{itemize}

\begin{table}[t!]
\centering
\small
\caption{\textsc{Characteristics of the Pilot Study Design}}
\vspace{-2.5ex}
\label{design:table}
\begin{tabular}{|c|c|p{0.7cm}|p{1.39cm}|p{1.39cm}|p{1.47cm}|} \hline
\textbf{Session} & \textbf{Case} & \textbf{Count} & \multicolumn{3}{c|}{\textbf{Subjects Group}}  \\ \hline
I  & A & S1-S56 & G1-2-6-7 \hspace{20 mm} (6 student) & G3-G5 \hspace{20 mm} (11 student) & G4 \hspace{20 mm} (10 student) \\ \hline
II & B  & S57-S112  & G8-11-13-14 \hspace{20 mm} (6 student) & G9-G10 \hspace{20 mm} (11 student) & G12 \hspace{20 mm} (10 student) \\ \hline
\end{tabular}
\vspace{-4.3ex}
\end{table}

\begin{figure*}[!ht]
\centerline{\includegraphics[height=1.5in,width=7.1in,angle=0]{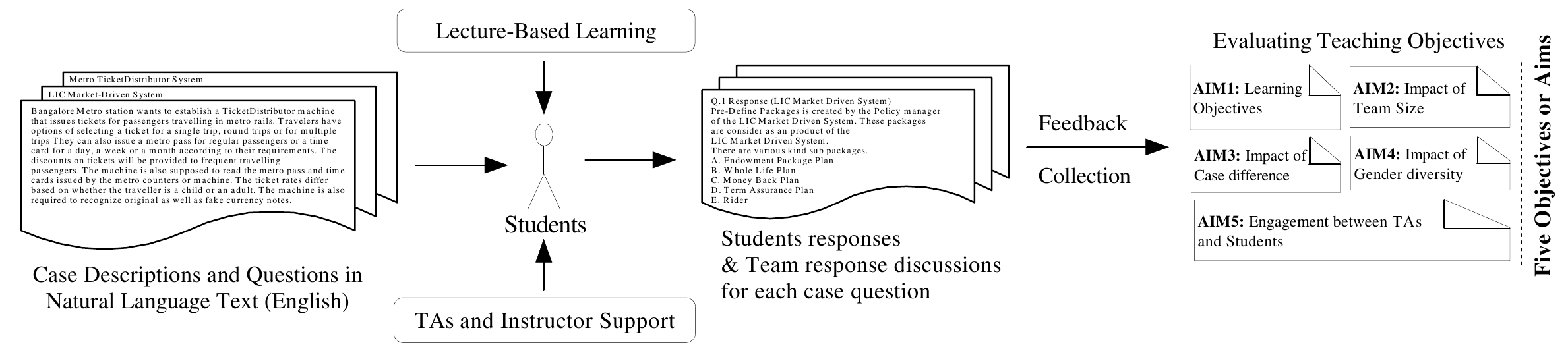}}
\vspace{-2.5ex}
\caption{Overview of the study for evaluating the effectiveness of CBL in students' learning}
\vspace{-2.5ex}
\label{fig:fig25}
\end{figure*}

\section{CBL Execution and Data Collection}
In this section, we provide details about the preparations done for CBL including students training, case descriptions, and CBL execution. The overview of the empirical study is shown in Figure~\ref{fig:fig25}. 

\subsection{Preparation and Subject Training}
Before commencing the CBL exercise, seven traditional lecture sessions were organized to familiarize students with the RE concepts. Additionally, a two-hour lab session was conducted to introduce them to the concepts of CBL through a class presentation, learning videos and demos. The course instructor planned to demonstrate CBL to TAs (two master students, one doctoral student, and one research fellow) such that the team (educators and TAs) could help students groups more efficiently in understanding the concepts.

\subsection{Execution}\label{execution}
The implementation of CBL was done in two different sessions each spanning two hours. In the first session, 56 students (7 groups) were assigned Case A (\textit{Metro Ticket Distributor System}), and demonstrated how the case-based exercises are required to be handled. The students groups were asked to start working on the case in the lab session itself, where TAs and the instructor were available to help them in analyzing and understanding the requirements specified in the case. The students were given one week time to solve the case (takeaway, not as a classroom exercise) such that students get sufficient time to analyze the case and associated multiple resources. After the case solving exercise was completed, a four-hour session was conducted for case discussions and analysis. Similarly, for Case B (\textit{LIC Market-Driven System}) the aforementioned process was followed for remaining 56 students. 

For each case, the study was formally carried out in three different phases, `Case Understanding', `Case Solving', `Case Discussions'. In \textit{Case Understanding Phase}, each team was allotted one of the two cases as shown in Table~\ref{design:table} and was asked to study the case carefully. In \textit{Case Solving Phase}, each student was asked to take the responsibility of one or more questions depending on the team size. However, the response to each question had to be finalized through team work. In our study, both RE cases consist of eight questions each, where Q1 and Q3 are more complex than the other six questions. The questions were randomly assigned to the students groups in such a way that the responsibility of each question among the members of the group was balanced. The rationale and details of each group type with reference to the assigned questions are as follows:

\begin{enumerate}
    \item For the group size of 11 students, each student was responsible for one question and two students were responsible for solving Q1 and Q3. And, one student was responsible (who is a team leader) for the additional question.
    \item For the group size of 10 students, each student was responsible for one question and two students were responsible for solving Q1 and Q3. Also, all students in the team were responsible for answering the additional question.
    \item For the group size of 6 students, each student was responsible for one question, and remaining three questions (two case questions (Q1, Q3) and one additional question) were assigned among six team members such that each additional question was assigned to different pairs of students in the group. 
\end{enumerate}

It can be observed that each student in groups of 6 has more responsibility than their classmates in the groups of 10 or 11. Before conducting the session, we asked students to volunteer and undertake this exercise in the group of 6. The rationale behind different group types was to investigate an ideal team-size required for conducting CBL methodology for better results. This rationale supported by the results, benefits, and limitations for choosing different group types was explained to students after the successful completion of CBL sessions. 

Finally, in \emph{Case Discussions Phase}, each student group was asked to submit their responses to each of the eight questions on the response sheet in digital format, and then present their responses to the case questions through a formal presentation. The students were allowed to access all types of reference materials. A targeted case discussion was carried out after each group's presentation for responses to the case questions. All responses to the case and survey questions were collected using Google Forms. 

The CBL exercise was assigned a weightage of 10\% among all different assignments for the Software Engineering course. The best team in a session was allotted 10 extra marks that would be added to their end-semester examination marks. This was done to incentivize them toward further participation. The overall exercise was evaluated and graded by one faculty member, and four TAs from SE domain. Also, the students' responses were evaluated and graded both at student-level and group-level in the class-session itself. When one student team was presenting their solutions, other groups were asked to assess them on the scale of 1-5 based on their understanding of the problem and responses to each question. Since peer-evaluation~\cite{Weaver1986} is an active learning technique that causes students to think, ask questions and respond, we adopted this strategy to increase reliability and validity of the results~\cite{rf:peerreview}. 

\section{Data Analysis and Results}
This section presents the analysis of the data collected during the CBL sessions, to evaluate the impact of CBL on perceived understanding of the RE concepts. 

\begin{table*}[t!]
  \centering
  \small
  \caption{\textsc{Survey questions grouped by the respective learning objectives [SA: Strongly Agree, A: Agree, DA: Disagree, SD: Strongly Disagree]}}
  \vspace{-2.5ex}
    \begin{tabular}{|l|c|p{7.4cm}|c|c|c|c|} \hline
    \textbf{Teaching Objectives} & \textbf{Q.No.} & \textbf{Questions} & \textbf{SA} & \textbf{A} & \textbf{DA} & \textbf{SD} \\ \hline
    Learning & Q1    & I feel the use of case was relevant in learning about RE concepts. & 44.50\% & 53.60\% & 1.80\% & 0 \\ \hline
    Learning & Q2    & The case allowed for a deeper understanding of RE concepts. & 47.30\% & 48.20\% & 4.50\% & 0 \\ \hline
    Learning & Q3    & The case will help me to retain the different aspects of Requirements Engineering better. & 47.30\% & 49.10\% & 3.60\% & 0 \\ \hline
    Critical Thinking & Q4    & The case allowed me to view an issue from multiple perspectives. & 55.50\% & 42.70\% & 1.80\% & 0 \\ \hline
    Critical Thinking & Q5    & The case was helpful in synthesizing ideas and information presented in course. & 32.70\% & 56.40\% & 10.90\% & 0 \\ \hline
    Critical Thinking & Q6    & The class added a lot of realism to class. & 32.10\% & 57.80\% & 10.10\% & 0 \\ \hline
    Engagement & Q7    & I was more engaged in class when using the case. & 32.70\% & 52.70\% & 14.50\% & 0 \\ \hline
    Engagement & Q8    & The case discussion increased my interest in learning about Requirements Engineering. & 42.70\% & 49.10\% & 8.20\% & 0 \\ \hline
    Communication skills & Q9    & The case discussion strengthened my communication skills to speak in front of the audience. & 53.60\% & 42.80\% & 3.60\% & 0 \\ \hline
    Team work & Q10   & The case discussion increased my confidence to work in a team. & 63.60\% & 33.60\% & 2.70\% & 0 \\ \hline
    \end{tabular}%
  \label{tab:Survey Questions}
  \vspace{-1.6ex}
\end{table*}%

\subsection{Data Validation}
The collected data was analyzed to identify discrepancies in the students' responses. As it turned out, one student from group G1 submitted two different responses for the feedback questions and one student from group G8 did not attend the CBL implementation sessions. So the responses of these two students were eliminated from our analysis, hence a total of 110 data points (i.e., 74 (66\%) males and 36 (34\%) females) were analyzed to draw inferences. 

\subsection{AIM1: Analysis on Learning Objectives}
Table~\ref{tab:Survey Questions} shows the list of survey questions and students' responses to the CBL learning outcomes with regard to their learning, critical thinking, engagement, teamwork and communication skills. The responses were collected for each of the CBL learning outcomes in terms of Strongly Agree (SA), Agree (A), Disagree (DA), and Strongly Disagree (SD). We used the same CBL survey questions as utilized by Saini et al.~\cite{veena2017} and Kundra et al.~\cite{kundra2016} for assessing the efficacy of CBL. 

From Table~\ref{tab:Survey Questions}, it can be inferred that 98.1\% of total number of students agreed (SA + A) that the cases were relevant in learning various RE concepts (Q1), and 95.5\% of the students suggested that the cases allowed for a deeper understanding of case concepts (Q2). Almost 14\% of total students don't think that they were more engaged in class when working with the cases (Q7). One possible reason for this might be the fact that the CBL sessions were conducted as takeaway exercises. Collectively, more than 90\% of the students felt that the cases allowed them to view the problems from multiple perspectives (Q4), helped in synthesizing ideas (Q5), and added realism in the class (Q6). A significant 97.2\% students agreed that the case discussions increased their confidence to work in teams, while 96.4\% believed that it strengthened their communication skills. In summary, these strong agreements provide positive indications about the CBL exercises conducted in this work. 

\subsection{AIM2: Impact of Team Size on Learning}
The relationship between team size and productivity in a specific environment is a question of investigation in SE domain.  However, no theoretical arguments and empirical evidences are available in favor of either larger or smaller teams, hence it is hard to generalize an effective team size~\cite{mao2016experimental}. CBL exercises involve collaboration of student teams, where team members demonstrate and utilize their varied skills in complementary roles toward solving the case. Here, we intend to investigate the impact of two different team sizes on students learning. Table~\ref{tab:team-size} shows the team-wise percentages of SA, A, D, and SD against the teaching objectives to investigate the impact of team size on students' learning.

\begin{table}[!ht]
  \centering
  \small
  \caption{\textsc{Percentage of SA, A, D, SD sliced by team size }}
  \vspace{-2.5ex}
    \begin{tabular}{|c|c|c|c|c|c|c|c|c|} \hline
    \textbf{Q.No.} & \multicolumn{4}{c|}{\textbf{Small Group (5-6)}} & \multicolumn{4}{c|}{\textbf{Large Group (10-11)}} \\ 
\cline{2-9}    \multicolumn{1}{|c|}{} & \multicolumn{1}{c|}{\textbf{SA\%}} & \multicolumn{1}{c|}{\textbf{A\%}} & \multicolumn{1}{c|}{\textbf{DA\%}} & \multicolumn{1}{c|}{\textbf{SD\%}} & \multicolumn{1}{c|}{\textbf{SA\%}} & \multicolumn{1}{c|}{\textbf{A\%}} & \multicolumn{1}{c|}{\textbf{DA\%}} & \multicolumn{1}{c|}{\textbf{SD\%}} \\ \hline
    \textbf{Q1} & 40.0    & 57.8  & 2.2   & 0     & 47.0    & 51.5  & 1.5   & 0 \\ \hline
    \textbf{Q2} & 42.2  & 48.9  & 8.9   & 0     & 50.0    & 48.5  & 1.5   & 0 \\ \hline
    \textbf{Q3} & 35.6  & 60.0   & 4.4   & 0     & 53.0    & 40.9  & 6.1   & 0 \\ \hline
    \textbf{Q4} & 57.8  & 37.8  & 4.4   & 0     & 45.4  & 54.5  & 0     & 0 \\ \hline
    \textbf{Q5} & 35.6  & 55.6  & 8.9   & 0     & 30.3  & 59.1  & 10.6  & 0 \\ \hline
    \textbf{Q6} & 31.1  & 57.8  & 11.1  & 0     & 31.8  & 59.1  & 9.1   & 0 \\ \hline
    \textbf{Q7} & 33.3  & 53.3  & 13.3  & 0     & 31.8  & 53.0    & 15.1  & 0 \\ \hline
    \textbf{Q8} & 46.7  & 46.7  & 6.7   & 0     & 42.4  & 48.5  & 9.1   & 0 \\ \hline
    \textbf{Q9} & 57.8  & 35.6  & 6.7   & 0     & 51.5    & 45.5  & 0     & 3 \\ \hline
    \textbf{Q10} & 60.0    & 35.6  & 4.4   & 0     & 65.2  & 33.3  & 1.5   & 0 \\ \hline
    \end{tabular}%
  \label{tab:team-size}%
  \vspace{-1.6ex}
\end{table}%

In order to analyze the data for responding to AIM2, we framed an extended study question or a sub-aim as ``\textit{Are there any significant differences between the students' responses of smaller and larger groups?}'', and performed the t-test for the respective `agree' and `disagree' percentages. Our null hypothesis is that there is no relationship between the team size and the student learning during CBL execution. The alternate hypothesis is that team size influences student learning. The t-test results show a p-value of 0.2291 at a significance level $\alpha$ = 0.05. Based on this, we fail to reject the null hypothesis, and conclude that the size of the team does not have any effect on student learning. 

To further investigate the impact of team size based on case differences, each of the eight case questions supported by an additional question solved by the students were analyzed. The additional question was a brainstorming question which included the study of other related systems. This question was added to investigate the problem-solving skills when working in a team. Table~\ref{tab:evaluation-marks} shows the average marks awarded for each question by the evaluation team comprising four TAs, one instructor and various students groups. Each row represents the average evaluation marks awarded for each individual question to all 14 groups. Four small groups (G1, G2, G6, G7), three large groups (G3, G4, G5) were assigned Case A; and four small groups (G8, G11, G13, G14), 3 large groups (G9, G10, G12) were assigned Case B. The students from each group were evaluated and graded on the basis of nine (8 main and 1 additional) case questions, and each question carried 10 marks. So, the question set per case carried a total of 90 marks. The group who scored the highest marks for Case A was a larger group (G3) with a total of 70 marks, and the group which scored the highest marks for Case B was again a larger group (G9) with a total of 80 marks. Hence, the larger groups came out to be the best groups for both the cases. 

\begin{table}[!ht]
  \centering
  \small
  \caption{\textsc{Evaluation marks of the students responses submitted for case questions sliced by Case A \& B}}
  \vspace{-2.5ex}
    \begin{tabular}{|p{0.45cm}|p{0.16cm}|p{0.16cm}|p{0.16cm}|p{0.16cm}|p{0.16cm}|p{0.28cm}|p{0.28cm}|p{0.28cm}|p{0.16cm}|p{0.16cm}|p{0.16cm}|p{0.16cm}|p{0.28cm}|p{0.28cm}|} \hline
    & \multicolumn{8}{c|}{\textbf{Small Group}} & \multicolumn{6}{c|}{\textbf{Large Group}} \\  \cline{2-15}
    Q.No. & \multicolumn{4}{|c|}{\textbf{Case A}} & \multicolumn{4}{c|}{\textbf{Case B}} & \multicolumn{3}{|c|}{\textbf{Case A}} & \multicolumn{3}{c|}{\textbf{Case B}} \\  \cline{2-15}
      & G1 & G2 & G6 & G7 & G8 & G11 & G13 & G14 & G3 & G4 & G5 & G9 & G10 & G12 \\  \hline
    Q1    & 6     & 7     & 7     & 8     & 7     & 10    & 8     & 10    & 9     & 10    & 7     & 10    & 9     & 7 \\  \hline
    Q2    & 6     & 10    & 9     & 7     & 6     & 9     & 7     & 10    & 7     & 8     & 10    & 9     & 10    & 10 \\  \hline
    Q3    & 5     & 6     & 7     & 8     & 7     & 9     & 7     & 7     & 8     & 7     & 6     & 8     & 8     & 10 \\  \hline
    Q4    & 6     & 7     & 6     & 8     & 6     & 8     & 7     & 6     & 6     & 7     & 6     & 10    & 8     & 6 \\  \hline
    Q5    & 7     & 7     & 6     & 7     & 7     & 7     & 7     & 6     & 8     & 6     & 7     & 9     & 10    & 7 \\  \hline
    Q6    & 8     & 6     & 6     & 8     & 7     & 7     & 6     & 6     & 7     & 6     & 7     & 9     & 9     & 7 \\  \hline
    Q7    & 8     & 8     & 8     & 6     & 9     & 8     & 7     & 7     & 8     & 8     & 7     & 7     & 8     & 8 \\  \hline
    Q8    & 8     & 8     & 8     & 8     & 7     & 9     & 8     & 7     & 8     & 7     & 7     & 9     & 8     & 9 \\  \hline
    AQ    & 8     & 8     & 8     & 8     & 8     & 8     & 8     & 8     & 9     & 8     & 8     & 9     & 8     & 7 \\  \hline
    Total & 62    & 67    & 65    & 68    & 64    & 75    & 65    & 67    & \textbf{70} & 67    & 65    & \textbf{80} & 78    & 71 \\  \hline
    \end{tabular}%
  \label{tab:evaluation-marks}
  \vspace{-1.6ex}
\end{table}%

The detailed observation of the data shows that the difference of total marks between large groups and small groups for both the cases is negligible. For instance, G7 is a smaller group which was assigned Case A, and it scored a total of 68 marks. The total marks of G7 are closer to the total marks of G3, which is a larger group and also the best group of Case A with a total of 70 marks. This shows that both the smaller (G7) and the larger (G3) groups have a difference of two marks which is negligible. The average marks over all the smaller groups are 66.7, and the same awarded to all the larger groups are 71.8; which also shows not much of a difference. 

Arguably, we can say that on one hand each student of a larger group had a responsibility of a single case question, and hence she was able to properly utilize the given time in researching and identifying the best solution (for one question only). On the other hand, the students of a smaller group were responsible for answering multiple case questions and this might have resulted in scoring a lower average of 66.7. Another possible reason behind this lesser score could be the poor time management in finding answers to multiple questions. However, the overall analysis results suggest that the size of the team does not affect the students learning through CBL.

\begin{table}[!ht]
  \centering
  \small
  \caption{\textsc{Percentage of SA, A, D, SD sliced by Case}}
  \vspace{-2.5ex}
    \begin{tabular}{|c|c|c|c|c|c|c|c|c|} \hline
    \multicolumn{5}{|c|}{\textbf{Case A}} & \multicolumn{4}{c|}{\textbf{Case B}} \\ \hline
    \multicolumn{1}{|c|}{\textbf{Q.No}} & \multicolumn{1}{c|}{\textbf{SA\%}} & \multicolumn{1}{c|}{\textbf{A\%}} & \multicolumn{1}{c|}{\textbf{DA\%}} & \multicolumn{1}{c|}{\textbf{SD\%}} & \multicolumn{1}{c|}{\textbf{SA\%}} & \multicolumn{1}{c|}{\textbf{A\%}} & \multicolumn{1}{c|}{\textbf{DA\%}} & \multicolumn{1}{c|}{\textbf{SD\%}} \\ \hline
    \textbf{Q1} & 51.8  & 48.2  & 0     & 0  & 37.0    & 59.3  & 3.7   & 0 \\ \hline
    \textbf{Q2} & 46.4  & 53.6  & 0     & 0  & 48.1  & 42.6  & 9.3   & 0 \\ \hline
    \textbf{Q3} & 57.1  & 41.1  & 1.8   & 0  & 35.2  & 57.4  & 7.4   & 0 \\ \hline
    \textbf{Q4} & 60.7  & 37.5  & 1.8   & 0  & 51.9  & 46.3  & 1.9   & 0 \\ \hline
    \textbf{Q5} & 35.7  & 57.1  & 7.1   & 0  & 29.6  & 57.4  & 13.0    & 0 \\ \hline
    \textbf{Q6} & 30.4  & 58.9  & 10.7  & 0  & 33.3  & 57.4  & 9.3   & 0 \\ \hline
    \textbf{Q7} & 35.7  & 44.6  & 19.6  & 0  & 29.6  & 61.1  & 9.3   & 0 \\ \hline
    \textbf{Q8} & 42.9  & 48.2  & 8.9   & 0  & 42.6  & 50.0    & 7.4   & 0 \\ \hline
    \textbf{Q9} & 51.8  & 41.1  & 7.1   & 0  & 55.6  & 40.7  & 0     & 3.7 \\ \hline
    \textbf{Q10} & 69.6  & 26.8  & 3.6   & 0  & 57.4  & 40.7  & 1.9   & 0 \\ \hline
    \end{tabular}
  \label{tab:Case-based-Survey}
   \vspace{-3.4ex}
\end{table}

\subsection{AIM3: Case Difference}
Table~\ref{tab:Case-based-Survey} shows the case-wise percentage of SA, A, D, and SD against the teaching objectives represented by various questions to investigate the impact of case difference on students' learning. For this, we framed an extended study aim (expressed in the form of a question) as ``\textit{Are there any significant differences between the students' responses for Case A and Case B?}'', and performed the t-test for both agree and disagree percentages. Our null hypothesis is that there is no relationship between the case differences and the student learning during CBL execution. The alternate hypothesis is that the cases influence students' learning. Here, the null hypothesis (H$_{01}$) assumes no significant differences, whereas the alternative hypothesis (H$_{11}$) suggest the existence of significant differences between the students' responses for both cases. After performing the t-test for both agree and disagree percentages, we get a p-value of 0.8455, at a significance level $\alpha$ = 0.05. Thus, we fail to reject the null hypotheses. Therefore, our results suggest that the two cases are perceived as equally effective, and help students to achieve all five learning objectives.

\subsection{AIM4: Gender Diversity}
We performed a gender-specific analysis, and our extended study question for the same is, ``\textit{Are there any significant difference between the responses from male and female students?}''. Table~\ref{tab: Gender Survey Case} shows the percentages of male and female students agreeing/disagreeing (SA, A, D, SD) to each of the ten survey questions, respectively. Students were asked to report their gender in the survey form. Our null hypothesis is that there is no relationship between the gender and the student learning during CBL execution. We perform the t-test and find that p-value for agree\% is 0.710, at a significance level $\alpha$ = 0.05. Therefore, we fail to reject the null hypotheses. This means that there is no significant difference in the responses from male and female students. Based on these results (which are based on perceptions), we conclude that the difference in gender has no effect on the CBL outcome and it helps both male and female students equally in achieving various learning objectives.

\begin{table}[ht]
  \centering
  \small
  \caption{\textsc{Percentage of SA, A, D, and SD for the 10 questions sliced by Gender}}
  \vspace{-2.5ex}
    \begin{tabular}{|c|c|c|c|c|c|c|c|c|} \hline
    \multicolumn{5}{|c|}{\textbf{Male}} & \multicolumn{4}{c|}{\textbf{Female}} \\ \hline
    \multicolumn{1}{|c|}{\textbf{Q.No.}} & \multicolumn{1}{p{2.07em}|}{\textbf{SA\%}} & \multicolumn{1}{c|}{\textbf{A\%}} & \multicolumn{1}{c|}{\textbf{DA\%}} & \multicolumn{1}{c|}{\textbf{SD\%}} & \multicolumn{1}{c|}{\textbf{SA\%}} & \multicolumn{1}{c|}{\textbf{A\%}} & \multicolumn{1}{c|}{\textbf{DA\%}} & \multicolumn{1}{c|}{\textbf{SD\%}} \\ \hline
    \textbf{Q1} & 48.6  & 50.0    & 1.4   & 0  & 36.1  & 61.1    & 2.8   & 0 \\ \hline
    \textbf{Q2} & 48.6  & 51.4    & 0     & 0  & 44.4  & 41.7    & 13.9  & 0 \\ \hline
    \textbf{Q3} & 47.3  & 47.3    & 5.4   & 0  & 47.2  & 52.8    & 0     & 0 \\ \hline
    \textbf{Q4} & 56.8  & 41.9    & 1.4   & 0  & 52.8  & 44.4    & 2.8   & 0 \\ \hline
    \textbf{Q5} & 31.1  & 59.5    & 9.5   & 0  & 36.1  & 50.0    & 13.9  & 0 \\ \hline
    \textbf{Q6} & 33.8  & 54.0    & 12.2  & 0  & 30.6  & 63.9    & 5.6   & 0 \\ \hline
    \textbf{Q7} & 35.1  & 46.0    & 18.9  & 0  & 27.8  & 66.7    & 5.6   & 0 \\ \hline
    \textbf{Q8} & 39.2  & 51.4    & 9.5   & 0  & 50.0    & 44.4    & 5.6   & 0 \\ \hline
    \textbf{Q9} & 59.5  & 33.8    & 4.1   & 2.7  & 41.7  & 55.6    & 2.8   & 0 \\ \hline
    \textbf{Q10} & 63.5  & 35.1    & 1.4   & 0  & 63.9  & 30.6    & 5.6   & 0 \\ \hline
    \end{tabular}%
  \label{tab: Gender Survey Case}
   \vspace{-2.5ex}
\end{table}%

\subsection{AIM5: TAs' Perceptions with CBL}
All four TAs had no prior experience with CBL as a student or as a TA. Before the implementation of CBL sessions, TAs closely studied the research papers from SEABED, solved few cases, conducted discussions among themselves and with the instructor too. When the TAs were asked about which group type they found easy to manage or facilitate, two TAs believed that smaller groups are easier to manage; one TA stated that both group types are equally manageable, whereas one TA felt that facilitating larger groups was easier than managing the smaller groups. 
When the TAs were asked whether they found CBL for RE more useful than  traditional lecture-based learning for the students to grasp the underlying RE concepts, all TAs agreed that CBL is more useful than lecture-based learning for certain RE topics. They also stated that the total number of TAs who participated in the CBL exercise was sufficient for a class of 112 students. The CBL exercise involves self-evaluation and assessment which triggers new ideas and strategies that naturally vary from group to group, However, all TAs involved were overall satisfied with the students' responses to the case questions. After the first CBL exercise, TAs were feeling confident about assisting in another CBL session. On a question about the time spent, time allocation, and participation of each member of a group, TAs responded that (1) they spent equal time on each group, (2) the time allocated to each group for solving the case was sufficient, and (3) they found a balanced participation across members of the groups of both sizes. 

Another interesting aspect of this CBL exercise is assessing the impact of team size on CBL, hence we asked TAs to suggest what should be the better group size based on the TAs experiences with CBL sessions, including facilitating groups of different sizes. Two TAs admitted that smaller groups could be a better option, one TA suggested that any group size can be chosen, and one TA stated that larger groups should be chosen. The detailed TAs responses for the questionnaires can be downloaded online\footnote{https://sites.google.com/site/saurabhiiitdmj/resources/TA\_responses.pdf}. Overall, the TAs responses and their stated experiences showed that the CBL results in an increased TA involvement and satisfaction, along with high engagement between the organizing team (mainly TAs in this case) and students.

\section{Challenges and Recommendations}
In this section, we share the practical challenges encountered while exercising the CBL methodology, along with our recommendations.

\subsection{Challenges Faced}
While defining a case several points must be kept in mind: (1) a case must not be too complex and should be understandable to the students; (2) writing a case would demand a sort of ``reverse'' engineering approach, i.e., how should we define the case so that it takes the students to multiple resources; and (3) the questions attached to a case must invoke students into exploring a variety of resources including books, websites, blogs, discussion forums (both developer and general). One of the major aims of CBL is to send the students on a quest for the best (or the most appropriate) solutions. These solutions originated from different student groups form a common multi-perspective representation against each case question. This scenario should then trigger a discussion following the students' presentations. 

Another aspect of the CBL is teaching the concepts modeled in the cases. CBL is a active learning and teaching methodology, and the concepts have to be taught using traditional lecture-based teaching (i.e., classroom teaching). Prior to conducting the CBL sessions, the team must brainstorm on which topics are easier to be taught using traditional lecture-based learning than CBL, and hence should target to identify the concepts that can be better taught using CBL; in turn modeling such topics into the case descriptions. This would help students to concentrate only on those concepts to solve the given case. In this study, two cases were designed to teach RE concepts like requirement elicitation techniques, requirement analysis, requirement prioritization techniques, concept mapping, use cases and user stories. 
Based on the evaluation of marks and students' responses, we observed that the students were able to grasp the RE concepts related to requirements elicitation, prioritization, and documentation techniques. However, students have some difficulty in applying the requirement analysis techniques, which suggest that a more extensive case based on the analysis techniques to be framed to enhance the students' learning. One case developed for some of selected RE concepts may not be able to provide learning to the students for all articulated concepts, hence separate cases may be framed for each concept or group of concepts, and make it a challenging task for investigation. 

Based on the feedback submitted by the students, it can be clearly interpreted that they found CBL an innovative and interesting technique in SE education, and asked to conduct more CBL sessions in other topics of SE too. They explicitly mentioned that this exercise helped them to solve the same problem from different perspectives, thereby improving their critical thinking skills and understanding of the RE concepts more effectively. The students found themselves engaged in a kind of research activity and they accessed several resources to serve the questions. Some of the students found problems in understanding the scenario modeled in the case, i.e., ambiguous requirements with respect to the cases. As the case questions involved forming assumptions and each group had its own unique set of assumptions, so this approach was less acceptable to some of the students. Some students suggested extending the time duration given to solve the cases, so that they could understand the cases well and find better solutions.

\subsection{Recommendations}
Teaching RE concepts using an active learning method requires the course instructor to motivate students with the benefits of the method. CBL is a different kind of teaching methodology and students experienced it for the first time, hence introductory sessions on CBL are needed to be conducted in order to acquaint them with this approach. These sessions should consist of CBL videos, case examples, demonstrations, and CBL's practical relevance. 

We also recommend that the CBL sessions can be conducted as the lab exercise because designing and implementing CBL sessions is time consuming. It involves arranging extra sessions, in addition to their regular class schedule used for introducing CBL concepts, for making them understand the cases and carrying out necessary discussions. Both the TAs and students were new to the CBL methodology. As the TAs were the facilitators of the group of students to understand CBL, it was important to introduce them with CBL concepts, motivation and implementation process prior to the execution. All the TAs were instructed and guided through a well-designed execution plan by the course instructor. They were also asked to refer several resources like YouTube videos related to CBL, research papers and cases available on SEABED platform. 

CBL focuses on the best solutions and helps students to get a feel of how the concepts can be applied in real projects. The students were advised to look for the `best possible solutions' relevant to the case. However, we found that many of the students groups tried to find the `right answers' instead of digging into identifying the `best solutions'. We observed this phenomenon at the time of presentations also that the students were more interested in knowing the right solutions rather than focusing on the best solutions along with the related approaches followed to reach those solutions. Hence, it is recommended that the students should be cautioned against this practice before the execution of CBL sessions. 

We experienced CBL with both group types, smaller and larger. As each student in the group is responsible for identifying the best possible solution, it is recommended that the number of questions in the case should be equal to the number of team members in a group. This would help students to take responsibility for exactly one question for which student gets sufficient time. We did not impose any time limit for the presentations and discussions. As a result, the presentation lengths varied between 15 minutes to 45 minutes. Hence, we recommend limiting the discussion time, otherwise presentations would become less interesting. Each member in the group should have a direct interaction among themselves such that they can experience a self-learning environment. After the presentation, a discussion session of about 10-15 minutes involving all other groups should be provided. Overall, based on our experiences with CBL, the authors recommend that the educators can choose CBL for teaching certain topics of RE or other SE concepts and achieving various learning objectives.

\section{Threats to Validity}

\emph{Construct validity:} A total of 14 groups of different sizes were created specifically to analyze the impact of team size on students learning. Each student team was created based on the CPI such that the average CPI among students groups remained same (see Table~\ref{tab:table7}). Also, before choosing the group type, the students were informed about the rationale behind this kind of assignment. Our study tends to evaluate the impact of group size on the students learning using CBL. However, the investigation was done only with two groups of students of different sizes, posing a potential threat, which can be alleviated by experimenting with additional categories of different group sizes. Since our results are based on student surveys, we believe more such studies are needed to increase generalizability, credibility and reliability of the approach \cite{trochim2006nonprobability}. Another limitation of our work is that several components of our work is based on qualitative data and perceptions which is not generalizable as opposed to experimental studies \cite{trochim2006nonprobability}.

Another threat may be the distribution of case questions among the members of smaller and larger groups. The case questions were randomly assigned to the students, ensuring that one student in the group would not get the responsibility of multiple questions (by following the procedure mentioned in Section~\ref{execution}). The students were selected for the study from the number of students registered for the course, thus removing the possibility of self-selection. The subjects were provided sufficient teaching on RE concepts and rationale behind CBL methodology. We used unvalidated, single scales to measure perceived variables, so we have no way to evaluate construct validity.

\emph{Internal and Measurement Validity:}  The participants were not randomly assigned, so the study suffers from an unequal groups threat to internal validity. Students' perceptions of their own learning may not be accurate, compromising measurement validity. 

\emph{Conclusion and External Validity:} The participants may not be representative of all students in all requirements engineering courses, undermining external validity. There can be response bias, inflation of answers and the responses or outcome can be based on perceptions. 

\section{Conclusions}
This paper reports an exploratory pilot study on teaching various RE concepts as a part of a graduate level SE course, using Case Based Learning (CBL) method. 
The students' responses showed positive indication towards all five teaching objectives. More specific analysis of the students' responses revealed that the (1) difference in RE cases, (2) use of group types (small or large), and (3) gender diversity, do not affect the quality of solutions and students involvement in the CBL exercise. 

Overall, our results revealed that the CBL approach, with a well-designed case, is suitable for teaching and learning of RE concepts. Additionally, the TAs who are the facilitators of CBL sessions shared their experiences and recommendations, and suggested that CBL results in an increased engagement between the organizing team and the students. 

\bibliographystyle{ACM-Reference-Format}
\bibliography{sample-bibliography}

\balance
\end{document}